\documentclass[onecolumn]{article}
\pdfoutput=1 
\usepackage{cite}
\usepackage{graphicx}
\usepackage{amsmath}
\usepackage{authblk}
\usepackage[english]{babel}
\usepackage{blindtext}
\usepackage[utf8]{inputenc}
\usepackage[binary-units=true]{siunitx}
\usepackage{dirtytalk}
\usepackage{url}
\usepackage[dvipsnames]{xcolor}
\usepackage{subfig}
\usepackage{color,soul}
\usepackage{geometry}
\newcommand{\um}{\ensuremath{\mu\mathrm{m}}\xspace}
\newcommand{\cm}{\ensuremath{\mathrm{cm}}\xspace}
\newcommand{\mm}{\ensuremath{\mathrm{mm}}\xspace}
\newcommand{\ns}{\ensuremath{\mathrm{ns}}\xspace}
\newcommand{\Hz}{\ensuremath{\mathrm{Hz}}\xspace}

\newcommand{\MHz}{\ensuremath{\mathrm{M\Hz}}\xspace}
\newcommand{\V}{\ensuremath{\mathrm{V}}\xspace}
\newcommand{\Ar}{\ensuremath{\mathrm{Ar}}\xspace}
\newcommand{\COtwo}{\ensuremath{\mathrm{CO}_2}\xspace}
\newcommand{\vdrift}{\ensuremath{v_{drift}}\xspace}
\newcommand{\tzero}{\ensuremath{t_0}\xspace}

\newcommand{\tri}{\ensuremath{\mathrm{3/4}}\xspace}
\newcommand{\qua}{\ensuremath{\mathrm{4/4}}\xspace}
\newcommand{\mrad}{\ensuremath{\mathrm{mrad}}\xspace}

\title{\bf{Muon trigger with fast Neural Networks on FPGA, a demonstrator}}
\author[1]{M. Migliorini}
\author[1,2]{J. Pazzini}
\author[3]{A. Triossi}
\author[1, 2]{M. Zanetti}
\author[1]{A. Zucchetta}
\affil[1]{Istituto Nazionale di Fisica Nucleare (INFN) Sezione di Padova, Padova, Italy}
\affil[2]{Dipartimento di Fisica e Astronomia ``G.Galilei'', Università degli studi di Padova, Padova, Italy}
\affil[3]{Institut Pluridisciplinaire Hubert Curien, CNRS, Université de Strasbourg, Strasbourg, France}

\begin{document}
\date{}
\maketitle 

\begin{abstract}
The online reconstruction of muon tracks in High Energy Physics experiments is a highly demanding task, typically performed with programmable logic boards, such as FPGAs.
Complex analytical algorithms are executed in a quasi-real-time environment to identify, select and reconstruct local tracks in often noise-rich environments.
A novel approach to the generation of local triggers based on an hybrid combination of Artificial Neural Networks and analytical methods is proposed, targeting the muon reconstruction for drift tube detectors.
The proposed algorithm exploits Neural Networks to solve otherwise computationally expensive analytical tasks for the unique identification of coherent signals and the removal of the geometrical ambiguities.
The proposed approach is deployed on state-of-the-art FPGA and its performances are evaluated on simulation and on data collected from cosmic rays.
\end{abstract}

\section{Introduction}\label{sec:intro}

The current trigger systems of the LHC experiments are in several cases similar to their original version designed more than twenty years ago, based on technological standards available at the time. At the first trigger stage, L1, coarse granularity data get processed by custom electronics devised specifically to implement well (pre)defined tasks;
the accuracy with which the properties of the physics objects are assessed results thus rather poor, enough only to perform a first rough selection of the interesting physics events.
Major improvements made in the last years on several fronts can allow boosting the performance of the L1 systems, bringing them not far from what achieved by processing offline in software full granularity data.
Notably, flexible tools became available to develop complex Machine Learning models and deploy them on generic commercial FPGAs; sophisticated classification and reconstruction tasks can thus be executed in hardware taking advantage both of their remarkable performance and their tremendously short evaluation time.

At the Compact Muon Solenoid, CMS, the muon identification and reconstruction in the barrel region of the experiment  relies on the information provided by the Drift Tube chambers (DT) and by the Resistive Plate chambers (RPC); while the latter guarantee redundant timing information, the former can yield both precise tracking and reliable bunch crossing identification. Each DT chamber provides, by means of a Trigger Primitive Generator (TPG),  the local information at the basis of the L1 muon trigger chain, i.e. the track stubs (trigger primitives) consistent with the passage through the chamber of a charged particle. 

Differently with respect to the current setup, the front-end electronics being developed for the High Luminosity upgrade of the LHC (HL-LHC) will serve full granularity DT measurements to the TPG, which can then aim at producing high accuracy candidates. The HL-LHC poses however arduous challenges: firstly, the higher noise and higher detector occupancy due to the larger instantaneous luminosity; at the same time, the ageing of the detector will lead by then to an unavoidable loss of efficiency. Furthermore, in addition to the high accuracy for muons, the algorithm processing the DT data should be highly flexible and provide as little bias as possible to enable an efficient detection of hypothetical non-standard particles, like heavy stable charged particles or muons produced far from the collision point.
To address these challenges, a novel algorithm for the muon identification and track parameter estimation at L1 has been developed, based on the combination of a series of Artificial Neural Networks (ANNs) and analytic methods. 
In this work we describe in detail this algorithm and its implementation on a Xilinx Kintex Ultrascale FPGA; we also report on its performances  both on a dedicated simulation and on data collected by a muon telescope installed in the INFN National Laboratory of Legnaro (LNL). 
The latter will be described in the next Section~\ref{sec:setup}, the algorithm and its training will be the subject of Section~\ref{sec:algo}, whereas Section~\ref{sec:perf} will illustrate the performance.

\section{Experimental setup}\label{sec:setup}

\subsection{Drift Tube detector}\label{sec:telescope}

The algorithm presented in this work aims at processing the measurements of the Drift Tube detectors, one of the systems constituting the muon spectrometer of the CMS experiment~\cite{cit:CMS,cit:CMS_muTDR}.
The detectors installed in CMS consist of layers of tubes (or cells) mechanically arranged to provide a tri-dimensional estimation of the muon track: the basic component is made of 4 stacked layers of parallel tubes, assembled  in a ``staggered'' configuration by half of the cell width, to ensure time-tagging capability. 
In the following we will focus on such a configuration, henceforth referred to as``chamber''\footnote{In the context of the CMS experiment a group of 4 layers is referred to as ``superlayer'' whereas a ``chamber'' is the mechanical assembly of 2 superlayers measuring the muon trajectory in the bending plane and sometimes a third one rotated by $90^\circ$, measuring the trajectory in the orthogonal direction.}. 
A muon crossing the chamber's volume of a chamber induces up to four time-coherent signals (hits).
The cells have a transverse cross section of $42\times13\,\mm^2$ and are filled with a gas mixture of \Ar(85\,\%) and \COtwo(15\,\%), kept at atmospheric pressure. 
A multi-electrode design is employed to provide good stability of the drift field within each cell, to assume an almost constant drift velocity ($\vdrift\approx54\,\um/\ns$) of the electrons produced by the passage of a ionizing particle; in each cell the $50\,\um$ diameter anodic wire is kept at the nominal high voltage potential of $\mathrm{V}_\text{wire} = + 3600\,\V$, while the side walls cathodes are at $\mathrm{V}_\text{cathode} = - 1800\,\V$.
The shaping of the potential is provided with two additional electrodes, mounted on the top- and bottom-wall of the cell (parallel to the drift path), kept at the negative potential of $\mathrm{V}_\text{strip} = - 1200\,\V$.

The algorithm is tested on a muon telescope installed at the Legnaro National INFN Laboratories (LNL) and operated with cosmic-ray muons. 
Such telescope consists of 4 DT chambers, each with a roughly square shape of $70\times70\,\cm^2$ and made of layers of 16 cells, totaling 64 cells per chamber. The details of the telescope configuration will be presented in Section~\ref{sec:perf_lnl}.

\subsection{Electronics chain}\label{sec:electonics}

The charge deposited on each sensing wire is amplified, shaped, and discriminated against a given threshold by custom ASIC~\cite{cit:MAD} hosted on \mbox{front-end (FE)} electronics located within the gas volume of the chambers.
Signals passing the discriminator threshold are finally formed compatible with LVDS levels and sent to the \mbox{read-out (RO)} electronics. The RO is compliant with the electronics standards defined for HL-HLC (a.k.a. DT Phase-2 Upgrade~\cite{cit:phase2Elec}) since it employs the same communication protocol on the optical link and it adopts the same strategy for the time digitization of the FE signals.
Two Xilinx evaluation boards (VC707), hosting Virtex-7 FPGAs, receive the discriminated LVDS signals from the FE and implement the time-to-digital conversion (TDC), assigning a time-stamp to each firing channel.
For each VC707, a total of 138 TDCs are implemented exploiting the standard IOSERDES running at 1.2 Gsps.
A single VC707 is capable of performing the digitization of the entirety of the channels from two telescope chambers.
The additional TDCs available are used to digitize other external signals.
A pair of scintillator tiles equipped with photomultipliers are used to provide an unbiased external timing reference. The coincidence NIM signal of the two photomultipliers is reshaped into a LVDS and injected in one VC707 to be digitized.
The firmware (FW) of the Virtex-7 includes an instance of the latency-optimized GBTx-FPGA~\cite{cit:gbt_FPGA} to implement a serial link compatible with the GBTx chip.
The GBTx link is a high speed bidirectional optical link to be used simultaneously for data readout, timing control distribution, and the slow control and monitoring.
An external oscillator is used to generate a clock signal, distributed to both VC707 boards, and set to provide the $120~\MHz$ frequency required to drive the GBTx transceivers; such frequency is also up and down scaled to implement all the synchronous clocks needed by the TDC cores.
Most noticeably, to emulate the CERN Timing, Trigger and Control (TTC) Systems~\cite{cit:ttc} used to distribute the timing information for all the LHC experiments at CERN, a $40~\MHz$ clock is generated, compatible with the LHC bunch crossing (BX) frequency.
The TDC provides a time measurement of the rising edge of the input signal with respect to the rising edge of the $40~\MHz$ clock. A coarse counter then assigns the BX to the TDC measurement. The timestamp of the hits is therefore given with the granularity of 1/30\textsuperscript{th} of BX. 
All FE signals converted to TDC hits are finally fed into the GBTx core and sent to SFP+ transceivers for transmission over optical links.
The optical links from both VC707s inject the serialized data streams into the input links of a Xilinx KCU1500 board hosting a Kintex UltraScale FPGA (XCKU115), where the local trigger algorithm is implemented (see Section~\ref{sec:algo}). The KCU1500 allows to receive up to 8 GBTx links and can be plugged on a PCI Express bus of a data server for final transfer and storage of the data. 
Its FW processes the streams of the entire set of TDC hits using the reconstructed clock from the data itself. All the hits are eventually merged in a single stream and sent to a Direct Memory Access engine along with output of the local trigger algorithm.

\section{Algorithm overview} \label{sec:algo}

The aim of the TPG logic is to define the local position and crossing angle for muons traversing the detectors, which can be well approximated with a straight line path across the volume of a chamber.

For each DT cell the drift time, i.e. the time difference between the production of primary ionization and the signal deposition on the anode wire, can be translated to a spatial distance thanks to the \vdrift linearity. 
An inherent left-right ambiguity does however still persist, as no information on the side of the production is a priori available for each cell.
Moreover, in all most common applications of the DT chambers the absolute time at which the primary ionization occurs is not an information available from external measurements, and has to be identified in situ.

\begin{figure}[hbt]
    \centering
    \includegraphics[width=.6\textwidth]{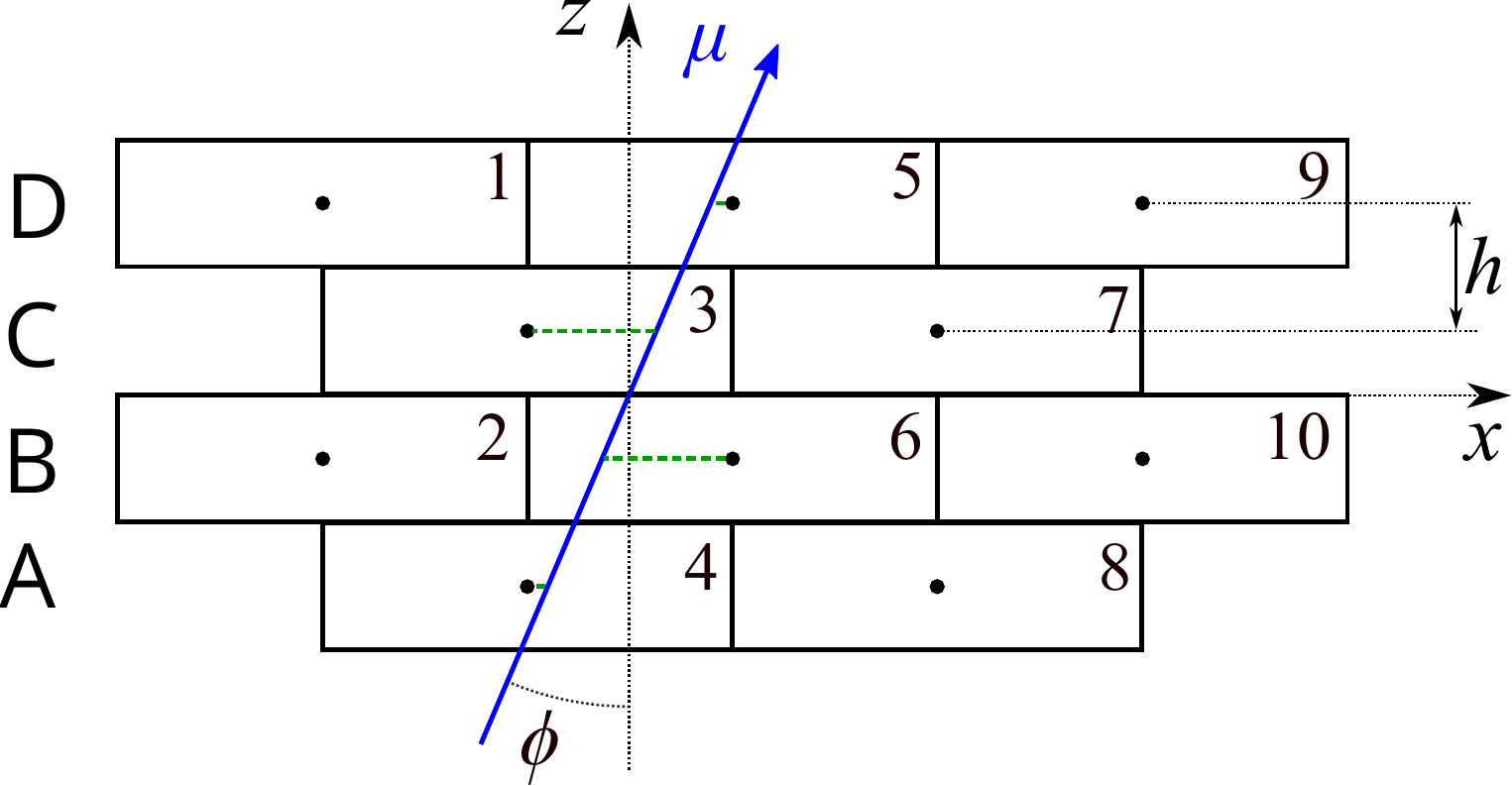}
    \caption{Schematic depiction of a muon cross one chamber. The active cells providing a signal are labeled 4-6-3-5.}
    \label{fig:mean-timer}
\end{figure}

Exploiting the generalized mean-timer technique\cite{cit:CMS_meantimer} the absolute time of passage of a particle can be extracted by combining the information of a set of 3 or more hits. The staggering of the cells in adjacent layers allows in fact to rely on the geometrical relationship existing among the drift paths in the crossed cells, extracting a common time pedestal (referred to as \tzero), and the local crossing angle. A small set of analytic relations can be identified depending on the geometrical pattern of the active cells and the left-right signal deposition, relating the signal collection time for each hit to the maximum drift time $T_\mathrm{max}$ allowed from the cell width. Using Figure~\ref{fig:mean-timer} as a reference, the mean-timer equation corresponding to the pattern of active cells 4-6-3 can be expressed as:
\begin{align}
    \tzero &= \frac{t_4+2t_6+t_3-2T_\mathrm{max}}{4}\\
    \phi &= \arctan\left(\frac{t_4-t_3}{2}\frac{\vdrift}{h}\right)
\end{align}

The left-right hit ambiguity prevents from identifying a single equation associated to the hit pattern, and typical TPG applications define logics based on iterative approaches or histogram-based majority to scan all possible combinations of patterns and equations.

With the increasing noise expected at HL-LHC, the number of spatial- and time-coherent signals is due to rise significantly.
Higher noise levels will in fact make the number of viable triplet combinations rise significantly, thus creating sizeable numbers of alternative mean-timer combinations to test in order to assess the time pedestal and track parameters.

The trigger has however to provide robust and reliable local muon identification and reconstruction within a maximum latency within a few microseconds, avoiding the generation of ghost-tracks as the result of spurious signal combinations.

While purely analytical approaches to the problem can be achieved, the combination of the low latency required to run this algorithm at the online level (within a finite time of about 1$~\mu s$ due to the L1 trigger limitations) and the robustness against the high noise rate is not granted.

ML methods and specifically various implementations of ANNs are commonly exploited to perform denoising tasks and pattern recognition. Among the features common to many ANN-based models, the fast evaluation time is one of their most appealing features for a TPG task. While analytical approaches to the noise rejection can take several iterations over a set of finite points, the evaluation time of an ANN is almost instantaneous, as it does not require any iteration and embeds the knowledge of the relations of the problem in the weights associated with the nonlinear activation functions. Once the best possible weights are defined on the basis of a training process, the evaluation of an ANN is comprised of a finite set of additions and multiplications, which can take of the order of tens of nanoseconds.

An alternative approach to the local trigger generation for the DT chambers is proposed, based on an hybrid approach: ANNs are used to perform noise rejection and pattern recognition on clusters of compatible TDC hits, followed by analytical relations for evaluating the time pedestal and the track parameters.

The goal of the ANNs is to solve the combinatorial prior to applying the mean-timer equations.
For each new set of TDC hits as input, the ANNs provides a unique combination of signals compatible with the passage of a muon, solving as well the left-right ambiguity for each TDC hit included in the pattern.
With a completely disambiguated set of hits as a result, there is only one single mean-timer equation associated to the pattern, and thus one single analytic relation is required to evaluate the time pedestal, and univocally locate the coordinates of the particle crossing the active volume of the detector.

\begin{figure}[htb]
    \centering
    \includegraphics[width=\textwidth]{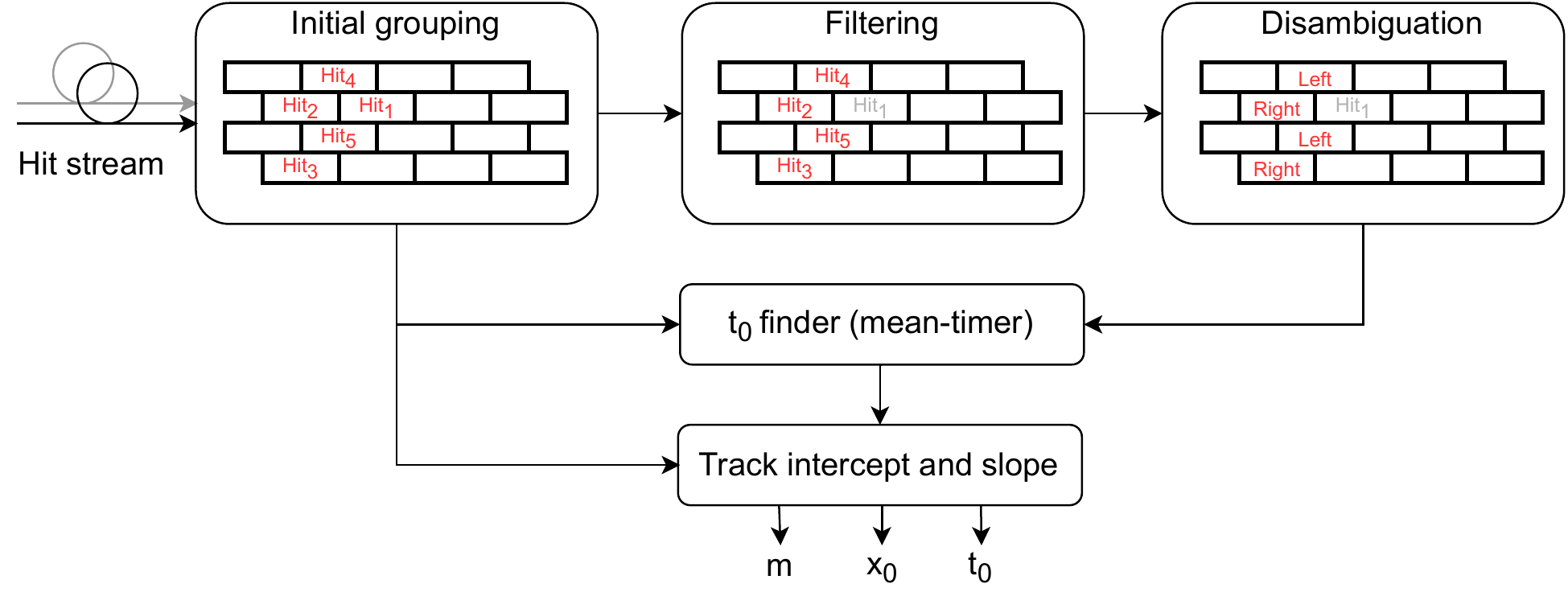}
    \caption{Schematic representation of the algorithm for each macro-cell. In the initial grouping hits are collected from the stream and positioned in the macro-cell. This is then fed to filtering and disambiguation blocks where noise is removed and the left-right ambiguity solved. This information is used to compute the muon crossing time and the track parameters.}
    \label{fig:diagram}
\end{figure}

In this section, the algorithm is described taking into consideration a single macro-cell, i.e., a group of $4\times4$ staggered cells, as depicted in Figure~\ref{fig:diagram}. The single macro-cell can be considered as an elementary building block for the proposed algorithm and it can be used to evaluate the performance of the method.
The macro-cell geometry is chosen to optimize the performance of the trigger algorithm under a number of circumstances, including both situations where the incoming particles are mainly produced from collisions (e.g., in test beams) and pointing toward the interaction spot, as well as applications where non-bunched particles are produced, as in the case of muon tomography with cosmic rays.
The algorithm can be divided in 5 sequential steps as illustrated in Figure \ref{fig:diagram} diagram. The first module, initial grouping, has the goal of collecting the hits from the input stream and organizing them in time-coherent sets for each macro-cell. The grouped hits are then fed to the two blocks containing neural networks. A first ANN is devoted to a filtering application, by classifying and selecting only hits compatible with genuine muons, and rejecting all spurious signals.
A second ANN performs a further disambiguation step, by targeting the identification of the side of the muon passage inside a cell with respect to the wire (referred to as hit \emph{laterality}), without the need to consider all possible combinations.\\
Out of the several hits possibly included in the original group, one single set of three or four filtered hits, already provided with the tentative laterality information, is extracted and passed to the following blocks, where fully analytical operations are performed, applying the appropriate mean-timer equation to evaluate the time pedestal. The \tzero information thus evaluated is then used in the last algorithm module to map hits in the coordinate space of the detector geometry, and finally perform a linear regression for obtaining the local track parameters.\\
In the next sections the modules are described in detail.  

\subsection{Initial grouping}\label{sec:algo_group}
\begin{figure}[htb]
    \centering
    \includegraphics[width=.8\textwidth]{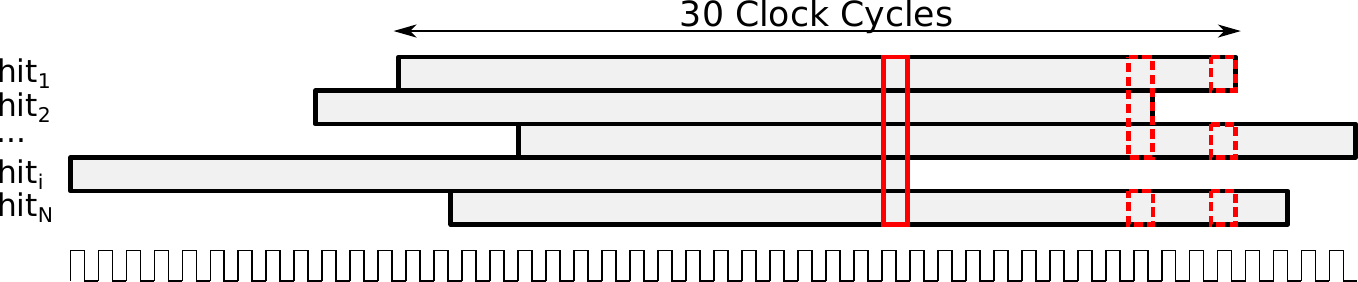}
    \caption{Illustrative depiction of the grouping and duplicate removal mechanism. Every hit is persisted for a 30 clock cycle long time window. When a hit ($hit_i$ in figure) reaches the end of the window, all the available hits are sent to the ANN filter block. Once a new hit ($hit_2$ in figure) reaches the end of the window, the set of available hits is compared with the previously processed one and discarded, as it is a sub-set.}
    \label{fig:grouping}
\end{figure}

This first module performs the collection of the hits deserialized from each of the GBTx lanes enabled on the KCU1500. Every hit encodes information about the timestamp associated with the TDC implementation on the VC707 and the channel, the latter corresponding to the physical wire collecting signals. It is worth stressing that the hits' order of arrival at the GBTx receiver links may be different from the one inherently due to their time tag, due to the funneling and multiplexing of the TDC blocks on each of the VC707 boards. This imply that the hit collection time window has to be set in such a way to be optimized to account the difference between the hit collection and arrival time.\\
Every hits is therefore persisted for a number of 30 clock cycles, where the clock is set at $40~\MHz$, to provide a window wide enough to include the maximum drift time inside a cell, of about 16 BXs, and the maximum expected multiplexing delay.\\
Once a hit reaches the end of its persistence, the current set of hits included in the time window is sent to the filtering block. This logic ensures that the hits produced by any given muon are processed together at least once. 
Additionally, a duplicate suppression step is implemented to discard multiple sub-patterns for which the same set of hits would be reprocessed several times in the following steps of ANN prediction. 
Given a set of hits $\{\text{hit}_i\}_{i \in \mathcal{S}}$ in a given time window, schematically represented in~Figure~\ref{fig:grouping}, once the first hit reaches the end of the persistency, all available hits are collected and passed to the filtering block downstream.
The filtering ANN will then proceed to identify a subpattern compatible with a genuine muon $\{\text{hit}_j\}_{j \in \mathcal{P}}$.
Once the following hit reaches the end of its time window, the new set $\mathcal{S'}$ of hits is compared with the previously identified one $\mathcal{S}$ to check for possible duplicates. If a new set of hits is observed, all hits will be moved downstream and the filtering ANN will be evaluated; otherwise, the hits are discarded as duplicates. 

\subsection{Filtering}\label{sec:algo_filter}

The filtering module aims at eliminating hits associated with noise while retaining only signals compatible with the passage of a muon. At most one hit is retained for each layer of the macro-cell, as multiple hits from the same layer are geometrically incompatible with a single muon. The method thus aims at identifying triplets (\tri) and quadruplets (\qua) of valid hits, i.e., three or four hits in different layers compatible both in time and space: hits from a genuine muon are related by the time constraint provided by the speed of signal collection, and the geometrical pattern of the traversed cells.\\
This part of the TPG algorithm is implemented with a ANN taking as input the coarse-grain BX timestamp of every hit, one per channel, and returning a binary mask identifying whether or not the hit is associated to noise. The ANN structure is a simple feed-forward neural network with 20 hidden neurons between the input and output layers, with 16 units each. To reduce the resource utilization on the board, the weights of the network are quantized at 6 bits. In this way, the model uses only LUTs (6K) and no DPS blocks, which are available in a smaller number on the board. To further reduce the resource utilization, the model has been pruned during the training, i.e., some connections between neurons are severed to reach a sparsity of $60\,\%$. The filtering ANN evaluation time is 2 clock cycles, with a $40\,\MHz$ clock. Since the forward pass performed for the evaluation of the network involves only operations in two consecutive layers, the execution can be pipe-lined to provide one new input to the network per each clock, thus increasing the throughput with respect to algorithms requiring multiple iterations. \\
If the outcome of the filtering network identifies at least three valid hits compatible with a muon, the predicted mask is used to select the hits that will be sent to the next module. Otherwise, the hits are marked as noise and discarded.

\subsection{Disambiguation}\label{sec:algo_disamb}

One of the most arduous pieces of information to be extracted from the list of timestamps provided by a set of hits is to identify the side of passage of the hit with respect to the wire. Even under noise-free assumptions (or after the aforementioned filtering step), the inherently ambiguous time-to-space mapping gives rise to a sizeable combinatorial: e.g. for any \qua signal the four available timestamps can be aprioristically combined in $2^4$ combinations. 
The disambiguation block has therefore the goal of resolving the combinatorial without any iterative procedure. This step is once more performed with a feed-forward neural network. The ANN model takes as input a \tri or \qua signal, one hit per layer, as identified by the filtering block. Every hit is characterized by its timestamp and the wire number within the layer $\{(\mathrm{BX}, \mathrm{Wire})_i\}_{i=1,\dots, 4}$. Each input pair is classified as passing on the left of the wire, on the right of the wire, or as an empty/noise channel. 
The model is allowed to identify hits as noise also at this stage, a posteriori of the filtering block, as a further attempt to recover those hits mis-identified as signals from the previous ANN step.\\
As for the filtering network, this model consists of one hidden layer with 20 hidden neurons. The weights have been quantized during the training procedure at 6 bits. Moreover, a pruning of the model during the training procedure was used to obtain a sparsity of $50\%$. Also in this case the model uses only LUTs (5K) and no DPS blocks. The latency of the disambiguation ANN block is measured at 3 cycles at 
$40\,\MHz$ clock, while the execution can also in this case be pile-lined allowing an increased throughput.

\subsection{Time pedestal finder}\label{sec:algo_time}

After the disambiguation step, only groups of three or four hits associated with the passage of a muon have been selected, and the left and right ambiguity has been resolved. This information can thus be used to univocally and un-ambiguously identify the only mean-timer equation providing a solution for the specific case and geometry.
All 19 mean-timer equations have been implemented in FW and they are mapped to the roughly 200 valid patterns in the $4\times4$ macro-cell considered. 
The selected equation provides the muon crossing time $\tzero$, which is evaluated in full TDC precision (each TDC corresponding to about $0.83\;ns$). This allows to obtain a trigger primitives' time resolution of the order of a few $ns$, comparable to the actual offline reconstruction. 

\subsection{Track parameters estimation}\label{sec:algo_param}

\begin{figure}
    \centering
    \includegraphics[width=0.7\textwidth]{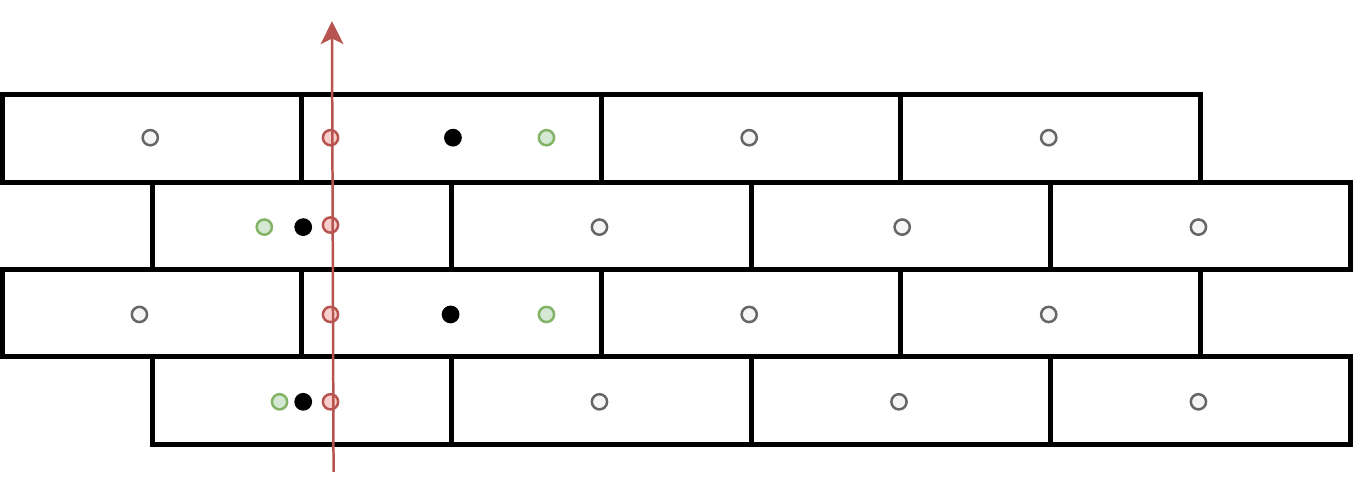}
    \caption{Example of a track in a macro-cell. The filled black circles represents the active wires, while the green and red markers depict the left-right ambiguity given the drift time. The red markers highlight the positions of the identified hit selected by the filtering and disambiguation blocks.}
    \label{fig:track_macro-cell}
\end{figure}

Once the \tzero has been evaluated for a triplet or a quadruplet, the hits' TDC signals are translated into the geometric coordinates of the expected points of production of the primary ionization, under the assumption of an approximately uniform $\vdrift$ across the entire DT cell.
In Figure \ref{fig:track_macro-cell} the red markers represent the position of the hits selected by the disambiguation network after the time pedestal has been subtracted. The trigger primitive track parameters, i.e., the angle with respect to the normal direction, and the crossing point in the middle layer of the macro-cell, are extracted by performing a least square linear regression considering the  \tzero-corrected hit positions. \\

\subsection{Simulation and Training}\label{sec:training}

\begin{figure}[htb]
    \centering
    \includegraphics[width=0.6\textwidth]{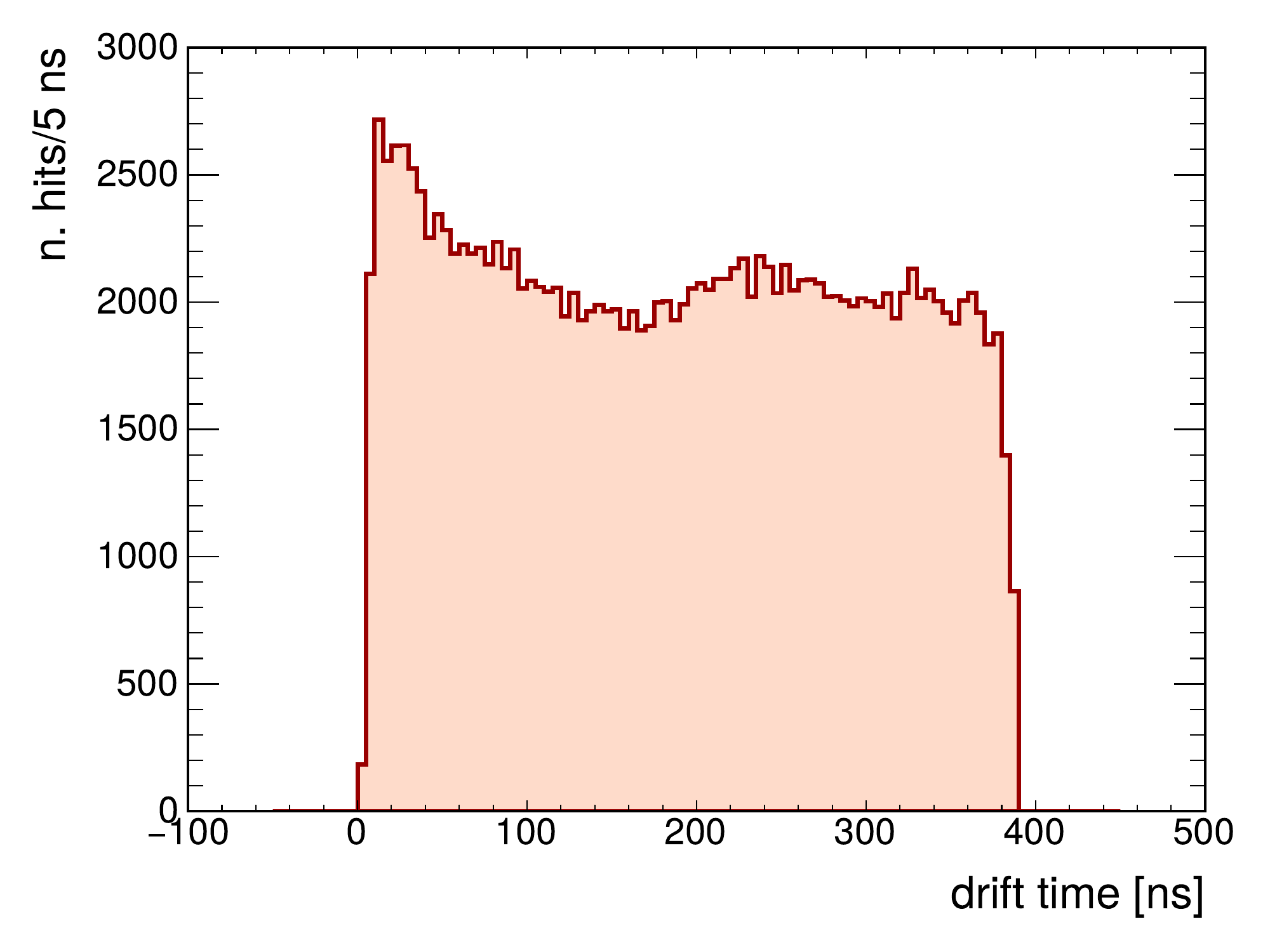}
    \caption{The pedestal-subtracted time distribution of hits produced in the private simulated sample (time-box). The shape of the time-box resemble very closely the one observed in data, indicating that detector behaviour is being correctly simulated.}
    \label{fig:simulation_timebox}
\end{figure}

The Machine Learning models used in this work fall under the supervised learning category, where the definition of the optimal weights for each ANN is carried out through a learning phase by exposing the ANNs to a set of examples (inputs) correctly labeled as the optimal desired output.
As the learning depends on data, the choice of the training dataset used is critical for the performance of the model. 
The dataset used in this work was produced from a private dedicated simulation of muons crossing the detectors under a broad range of conditions.
Using the studies performed on CMS Drift Tubes\cite{Puerta-Pelayo:904792,Benettoni:2007zz} the response of the DT detectors is modeled after the real data collected by the CMS collaboration to include effects such as electric field nonlinearity, cell inefficiencies, intrinsic cell position resolution, and detector ageing. 
Muons are generated according to a flat angular distribution in the range $[-45^{\circ},+45^{\circ}]$ with respect to the macro-cell view, with no tilt along the wire direction.
The broad angular spectrum of the generated muons does not reflect either the conditions muons are expected to be produced in collisions at colliders, nor the see-level spectrum of muons from cosmic rays, as in both cases the related probability density functions contribute to the enhancement of muons with small crossing angles. The choice is operated to expose the ANNs to all possible sets of angles and patterns.
The intrinsic resolution for the position measurement of the DT cells of $250\,\mu m$, as reported from~\cite{cit:CMS_performance}, is integrated into the simulation as a smearing of the hits' signal collection time to recreate a realistic sample.
Channel-by-channel efficiencies are also included by discarding random hits according to a Poissonian term, using as a reference the CMS DT single-cell efficiency estimation from~\cite{cit:CMS_performance}.
Additional hits, not associated with the generated muon, are also included in the simulation to emulate the presence of independent noise from a variety of sources, as, for instance, that induced by FE electronics. A $2\,\%$ Poissonian probability term is associated to each channel to give rise to a random hit, whose time is randomly generated.
All relevant features produced by the simulation, including the pedestal-subtracted time distribution computed from the simulated events (referred to as time-box) reported in Figure~\ref{fig:simulation_timebox}, are in excellent agreement with the ones obtained on real data. \\
Despite more extensive and comprehensive tools are available to perform more refined modeling of the detector's physics, including, for instance, the contributions of delta rays and material interaction, the simplified private simulations developed for the training of the ANNs provide a more flexible solution for training and testing purposes, allowing to generate $\mathcal{O}(10^6)$ events in a few minutes for every configuration under study, e.g. different rate of noise, or fraction of inefficient cells. \\
While the privately generated simulations excel in flexibility and provide a very close representation of the DT physics, future studies might be performed training the ML models on a more precise simulation; nonetheless, regardless of the level of detail included in the simulations, the advantage of an approach based on ANNs also resides in the fact that multiple variations of trained models will differ by a number of weights, and possibly some hyper-parameters of the model (such as the activation function used, or the amount of hidden neurons included in the model). Each new training can be deployed very efficiently on FW as an independent block, by updating the internal conditions of the model block without changing the up- and down-stream links and interfaces. \\

\begin{figure}[htb]
  \centering
  \subfloat[]{\includegraphics[width=0.4\textwidth]{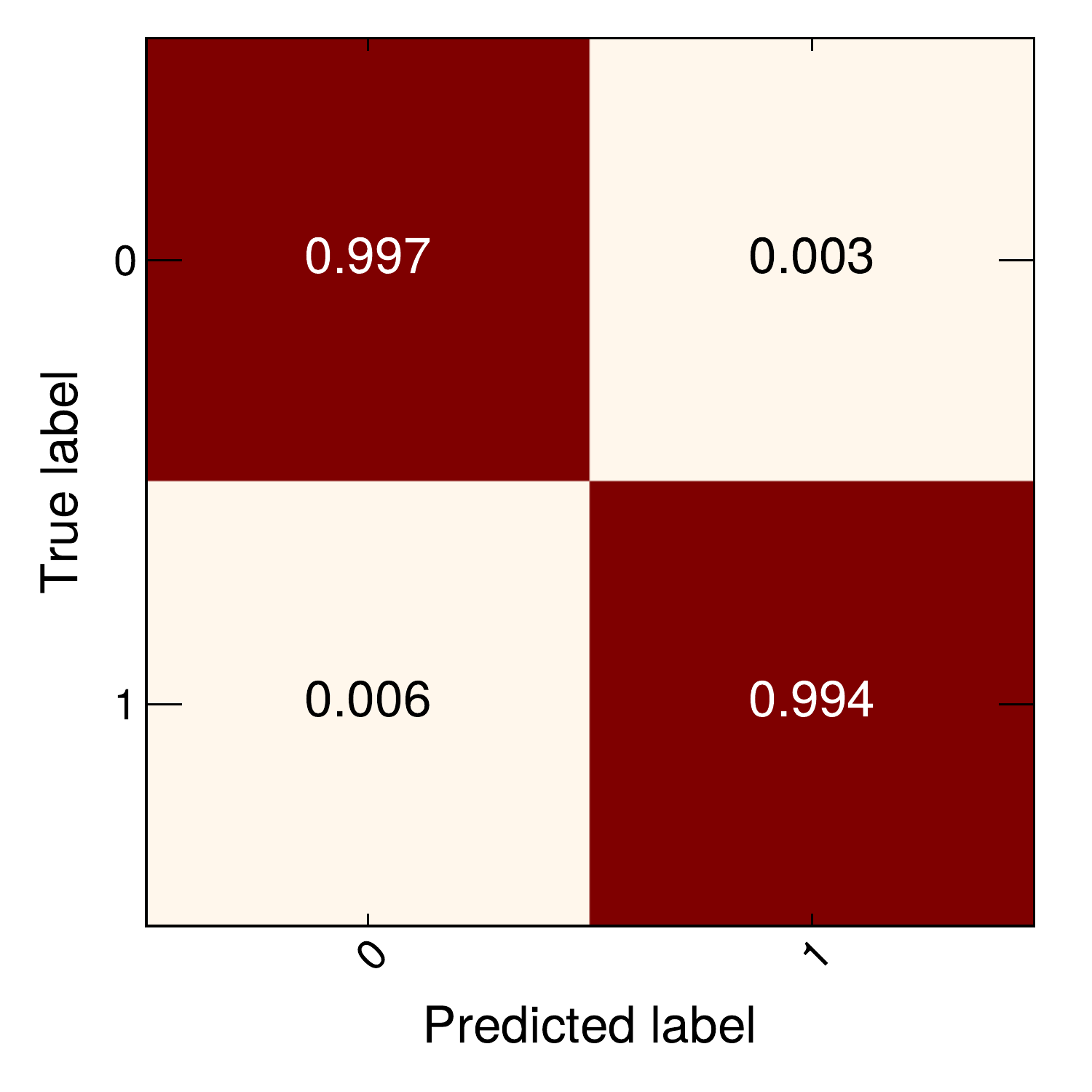}\label{fig:cm_filtering}}
  \hfill
  \subfloat[]{\includegraphics[width=0.4\textwidth]{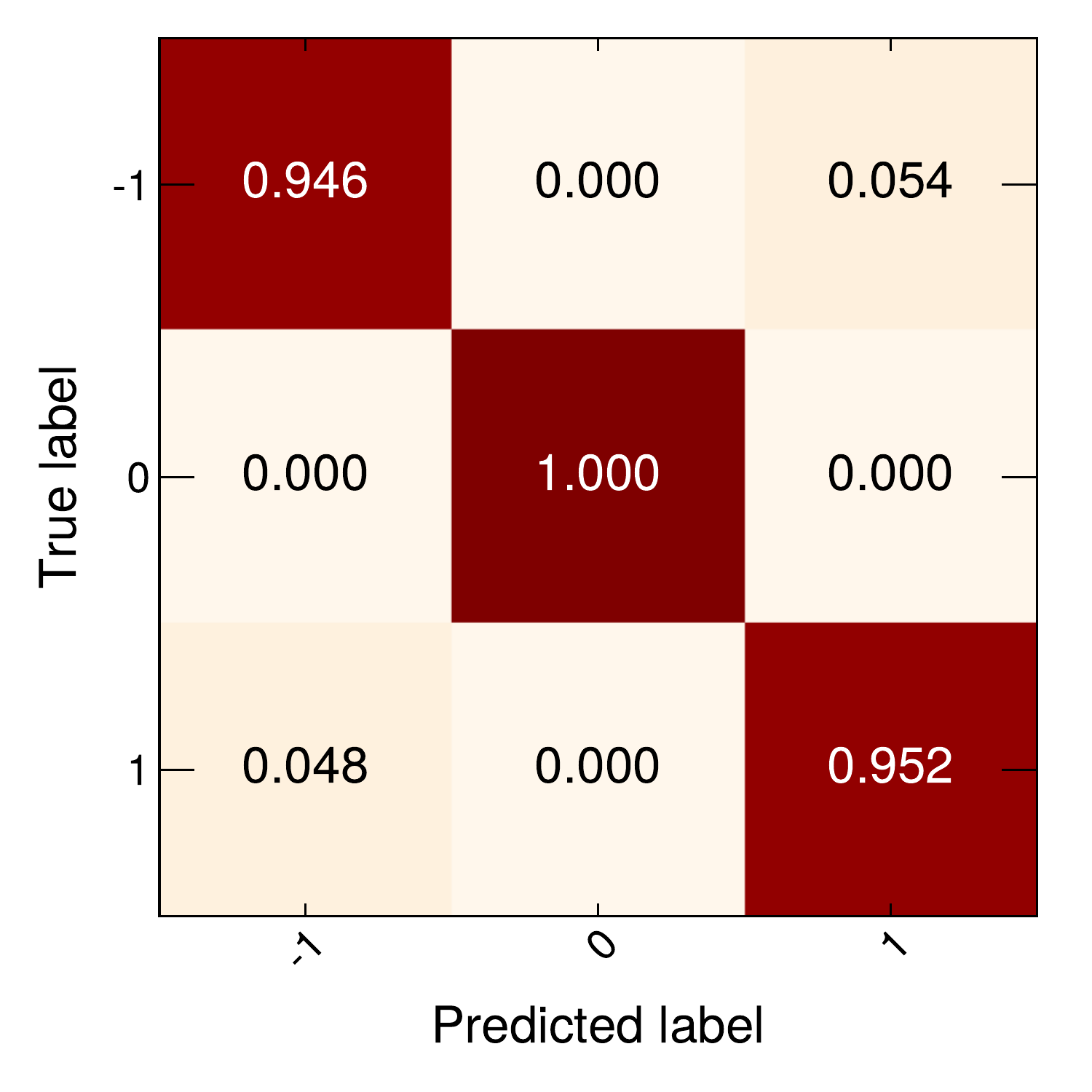}\label{fig:cm_disambiguation}}
  \caption{Confusion matrix for filtering network (a) and disambiguation network (b) evaluated on the simulated test dataset.}
\end{figure}

\noindent
The two ANN models have been trained using QKeras\cite{Coelho:2020zfu}, a library that can be used as a drop-in replacement of Keras\cite{chollet2015keras} to perform a quantization-aware training. To further reduce the area occupied by models on the FPGA, low magnitude weights are iteratively removed during the training procedure to obtain a sparse model\cite{2017arXiv171001878Z} with the trade-off of a limited reduction of the model performance. Models have been converted into HLS code using hls4ml\cite{Duarte:2018ite}, an open source library designed to provide a solution for the deployment of neural networks on FPGAs. The agreement between the Qkeras version of the model and the one implemented as HLS code blocks is excellent: provided the same input, the prediction of the two models is the same in the $99.9\%$ and $99.6\%$ of the cases for the filtering and disambiguation networks, respectively. \\
The performances of the models are evaluated on an independent test dataset by comparing the expected results known from the simulations with the output of the ANNs, as implemented in HLS. The results indicate an excellent performance for both networks, as can be seen from the confusion matrix in Figure \ref{fig:cm_filtering} and Figure \ref{fig:cm_disambiguation}, where the non-diagonal terms, indicative of wrongly classified examples, are found to be almost negligible. 

\section{Performance Evaluation}\label{sec:perf}

\subsection{Private simulation}\label{sec:perf_sim}

The performances of the proposed algorithm are tested on an independent set of events produced from the private simulation used to generate the ANNs training dataset.

\begin{figure}[htb]
  \centering
  \subfloat[]{\includegraphics[width=0.5\textwidth]{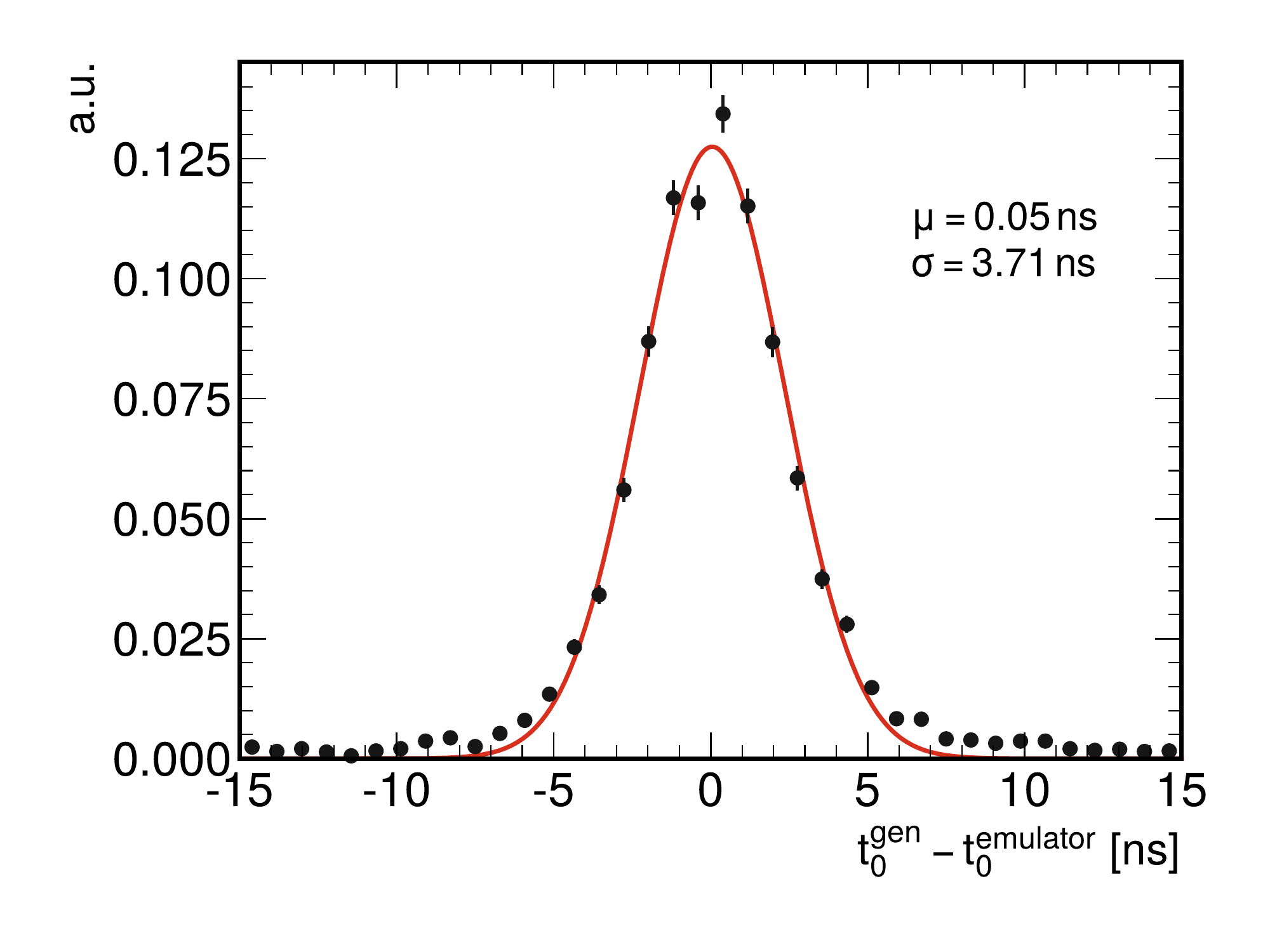}\label{fig:t0_emu_reco}}
  \hfill
  \subfloat[]{\includegraphics[width=0.5\textwidth]{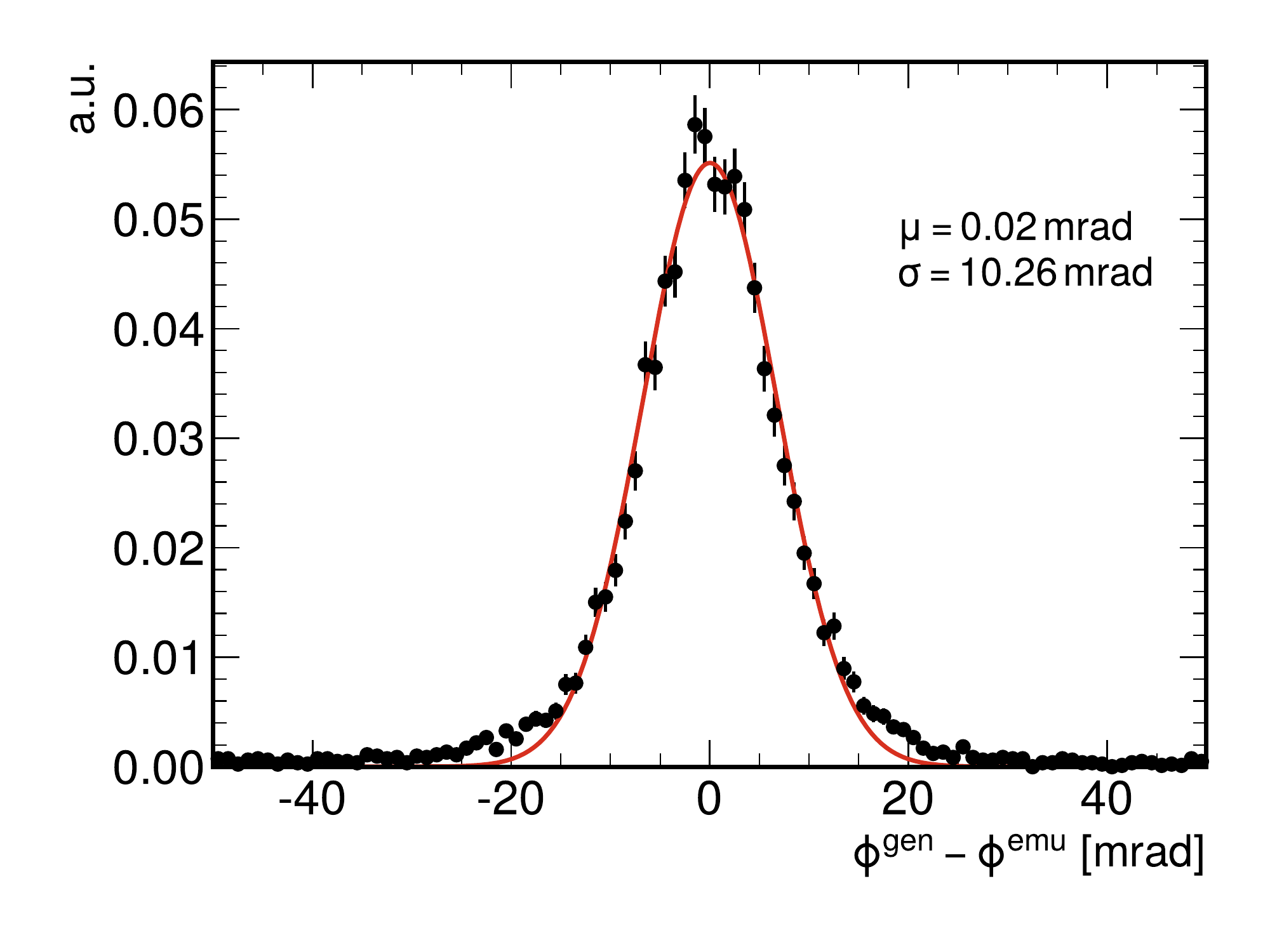}\label{fig:angle_emu_reco}}
  \caption{Time pedestal (a) and trigger primitive angle (b) resolutions evaluated on the simulated sample.}
\end{figure}

\begin{figure}[htb]
    \centering
    \includegraphics[width=0.5\textwidth]{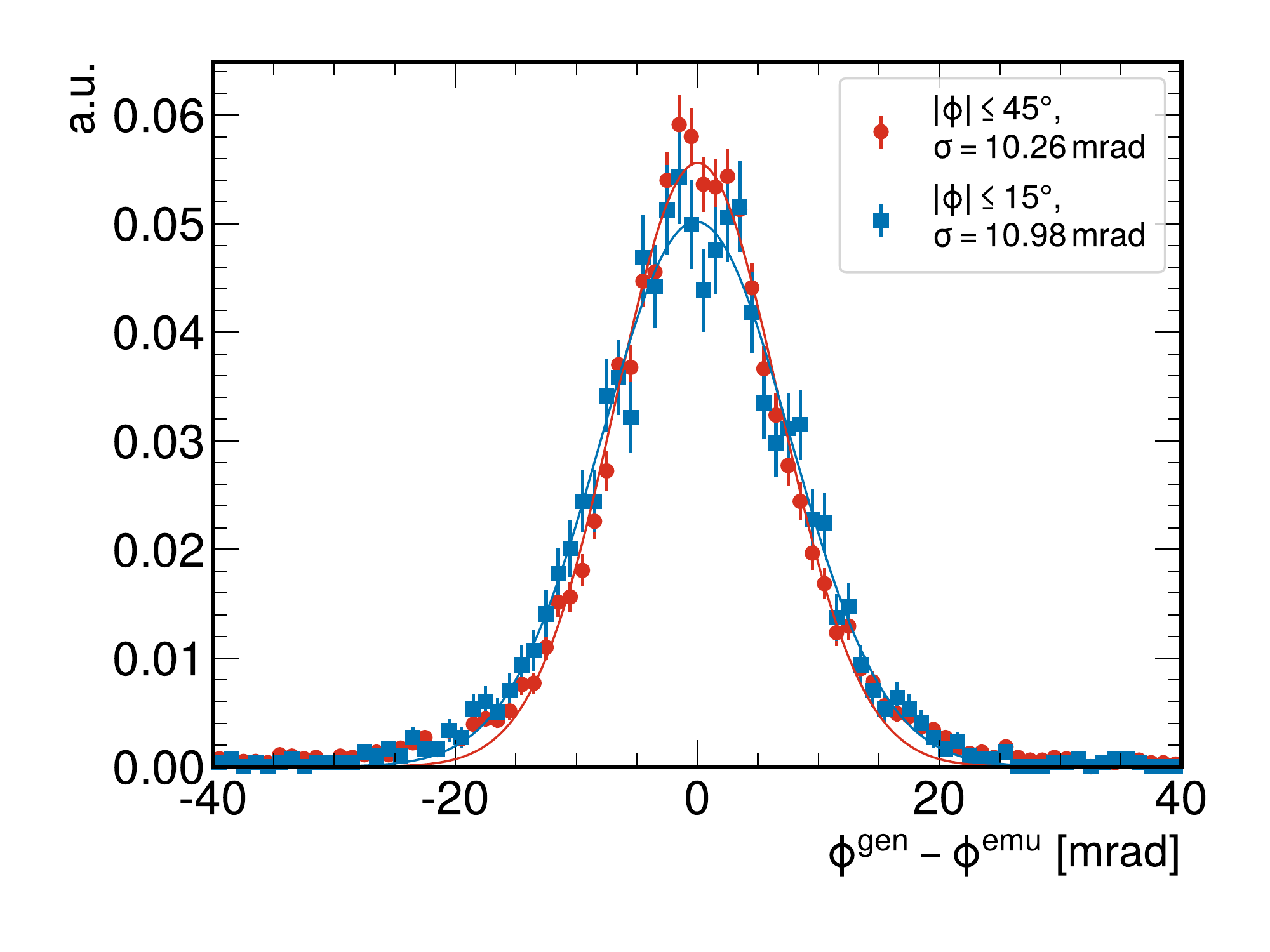}
    \caption{Trigger primitive angle resolution evaluated on the simulated sample for generated muons produced with a crossing angle $|\phi_{\mathrm{gen}}|<15^{\circ}$ (blue), and $|\phi_{\mathrm{gen}}|<45^{\circ}$ (red) respectively. The contributions are normalized to the same area.}
    \label{fig:local_angle_sep_sim}
\end{figure}

\begin{figure}[!h]
    \centering
    \includegraphics[width=0.5\textwidth]{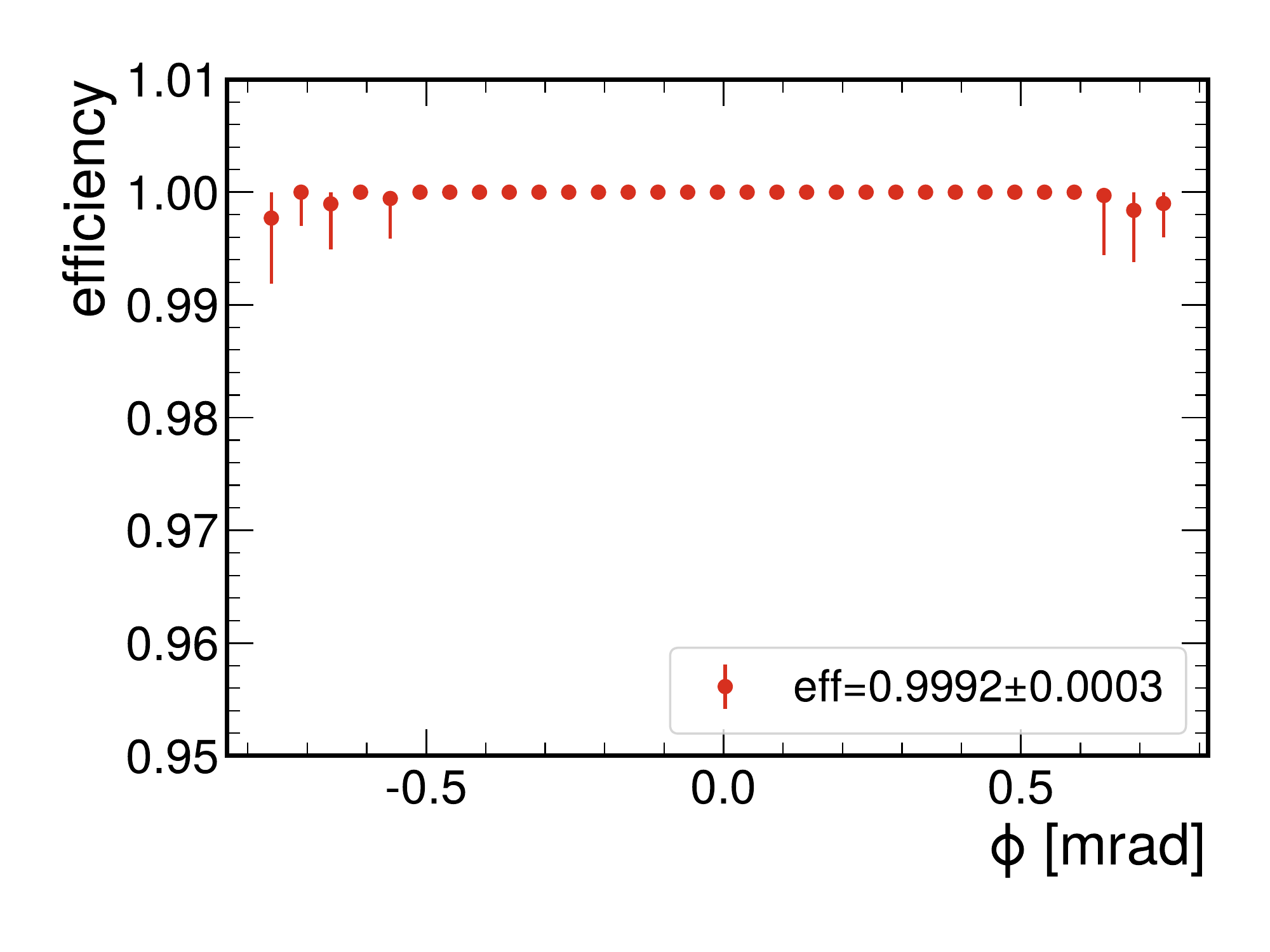}
    \caption{Trigger primitive efficiency of \qua signals as a function of the local angle $\phi_{\mathrm{gen}}$.}
    \label{fig:eff_angle_4hits}
\end{figure}

All hits associated with the generated muons, as well as the hits due to injected noise, are processed by a python-based software emulating the FW implementation of the algorithm.
A playback functionality of the software framework also allows to process the data collected by the hardware read-out, thus enabling a direct comparison of the SW and FW implementation of the algorithm.
An excellent agreement is observed between the two: the estimated emulator efficiency, i.e. the fraction of triggers produced by the FW for which a SW equivalent is produced, is $99.9\,\%$. The almost negligible fraction of muons not correctly identified by the emulator is due to differences between the QKeras ANN implementation included in the SW framework and the HLS implementation of the FW.
The time pedestal and angular track parameters produced by the FW implementation are emulated with excellent accuracy and precision. For all events included in this study the \tzero pedestal is emulated within $1\,\ns$ to the value produced by the FW implementation, compatible with the expected time estimation precision of 1 TDC count (where each TDC count corresponds to $0.83\,\ns$).
Minor discrepancies in the fit parameters are expected due to a different treatment of floating point precision in FW and SW: an RMS $0.27\,\mrad$ is observed for the distribution of the difference between the trigger primitive $\phi$ and the angle estimated in SW, a negligible contribution compared to the intrinsic detector angular resolution.
The SW emulator is used to evaluate the expected performance of the algorithm on the generated sample.
A time resolution of $\sigma_t=3.7\,\ns$ is estimated on the inclusive sample of generated muon, as reported in Figure~\ref{fig:t0_emu_reco}. The time resolution is compatible with the intrinsic spatial resolution of the cell ($250\,\mu m$) considering the uniform \vdrift assumption.
A $20\,\%$ fraction of the generated muons is simulated to produce 3 hits in independent layers as the result of a series of possible causes, e.g.: the simulated cell-by-cell inefficiency, the intrinsic macro-cell geometrical acceptance, or the simulation of a muon not releasing primary ionization due to its trajectory being compatible with the cells' side-walls.
The time resolution is expected to be almost independent of the number of hits composing the muon tracks.
This is a specific feature of the algorithm provided by the a-priori disambiguation of the hits, assigning the expected laterality to each signal. 
A mean-timer equation is therefore univocally assigned to any hit combination, hence a time pedestal can be extracted with similar resolution under both \tri or \qua signal assumptions.
A $\sigma_t^{\qua}=3.1\,\ns$ can be extracted for the \qua primitives, and a comparable, albeit slightly larger, resolution $\sigma_t^{\tri}=4.5\,\ns$ is estimated for the \tri. 
Figure~\ref{fig:angle_emu_reco} displays the inclusive angular resolution of the emulated trigger primitives, which is estimated as $\sigma_\phi=10.26\,\mrad$. The angular resolution reflects the additional degree of freedom in the least squares linear regression of the segment of \qua compared to \tri signals, resulting in a $\sigma_\phi^{\qua}=9.95\,\mrad$ and $\sigma_\phi^{\tri}=13.28\,\mrad$. 
An important figure of the performance of a trigger primitive algorithm is the uniformity of the angular resolution with the track's crossing angle.
This is reported in Figure~\ref{fig:local_angle_sep_sim} where the angular resolution evaluated on the inclusive sample ($|\phi_{\mathrm{gen}}|<45^{\circ}$) is compared with a subset of events with mostly perpendicular tracks ($|\phi_{\mathrm{gen}}|<15^{\circ}$).
The results show an almost flat $\sigma_\phi$ profile with $\phi_{\mathrm{gen}}$.
It is worth highlighting that the muons are generated under a flat angular probability distribution, hence populating uniformly both low- and high-angle fractions of the phase-space.
The efficiency of the TPG algorithm is defined as the fraction of generated muons for which a trigger primitive is provided by the emulator. An overall efficiency $\varepsilon=99.0\pm0.1\,\%$ is estimated from the whole simulated sample. The efficiency for the \qua signals is reported in Figure~\ref{fig:eff_angle_4hits} as a function of the muon angle, where a flat efficiency profile in the whole $[-45^{\circ},+45^{\circ}]$ range can be observed, with an average value of $\varepsilon^{\qua}=99.92\pm0.03\,\%$.

\subsection{Performance on data with cosmic rays}\label{sec:perf_lnl}

\begin{figure}[htb]
    \centering
    \includegraphics[width=0.45\textwidth]{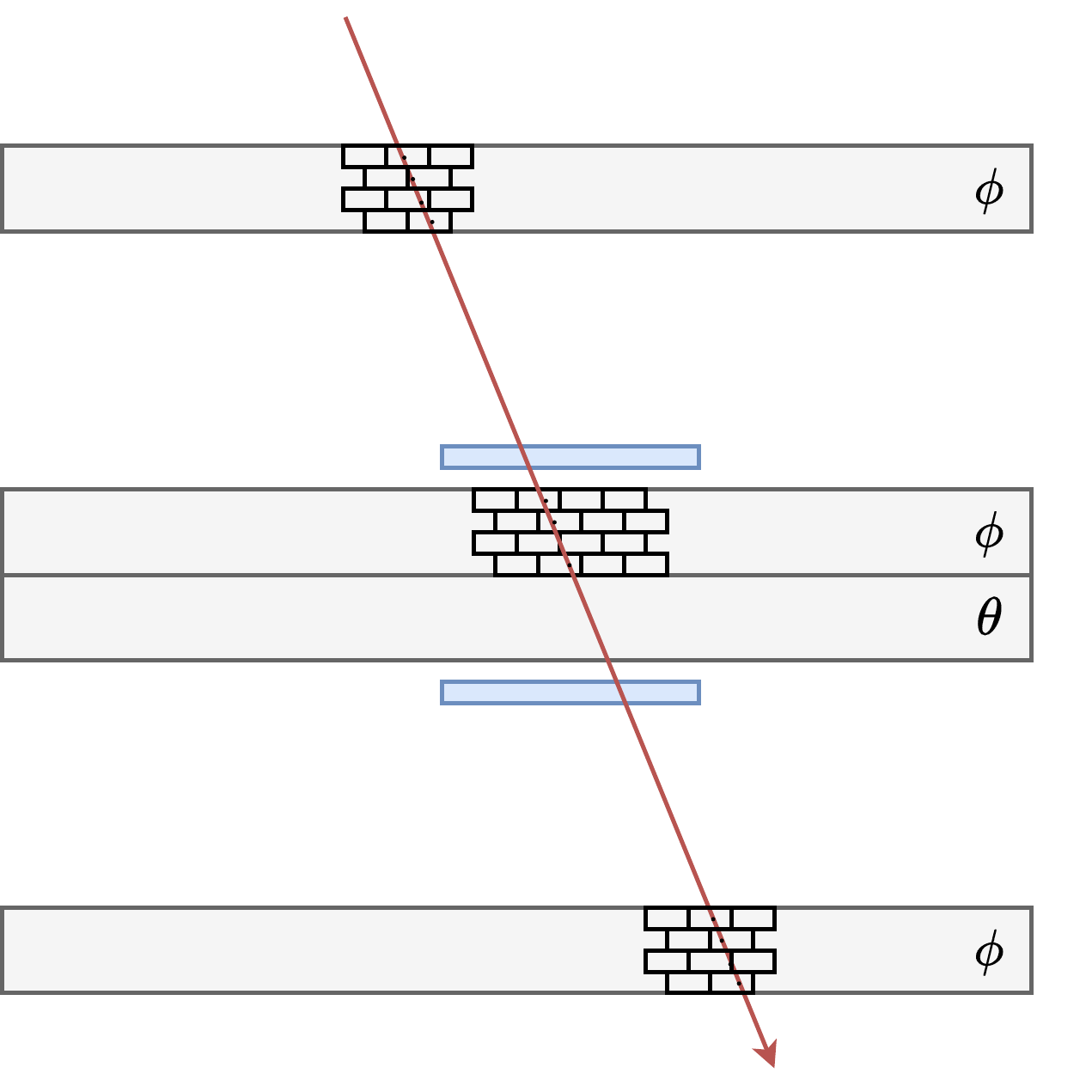}
    \caption{Simplified depiction of the LNL telescope geometry. The 4 mini-DT chambers are depicted as the gray boxes, marked with the corresponding $\phi$- or $\theta$-view. Two scintillator palettes, in light blue in the figure, are employed to provide an external time information.}
    \label{fig:telescope}
\end{figure}

The algorithm performances are tested on real cosmic ray data by deploying the FW implementation of the algorithm in the KCU1500 board of the LNL testbed, where 4 mini-DT chambers have been stacked in a telescope configuration.
The telescope setup is composed by 3 chambers in the same view (for consistency, referred to as $\phi$-view), and a fourth chamber providing the orthogonal view (referred to as $\theta$-view), to measure the complementary information.
The two most external $\phi$-chambers, installed roughly $80\,cm$ apart from both ends to the central pair of $\phi$- and $\theta$-chambers, provide a robust setup for tracking the passage of cosmic rays and interpolating the particle trajectory from the outer chambers to the central ones.
Trigger primitives are generated for a macro-cell of channels belonging to the mid-plane $\phi$-chamber. The KCU1500 board equipped with the FW implementation of the algorithm for the test macro-cell.
A schematic view of the geometry of the telescope is illustrated in Figure~\ref{fig:telescope}.

A pair of scintillator read out in coincidence embraces the test $\phi$-chamber providing an external timing reference.
The data acquisition system of the telescope allows to select all TDC hits and all trigger primitives produced, as well as the external scintillator coincidence signal, for storage and offline analysis.
The offline reconstruction of the muon tracks is performed selecting only those hits whose TDC timestamp is compatible with the external scintillator coincidence within the time window of the maximum allowed drift time ($\approx390\,\ns$). A constant time calibration, specific to each chamber, accounts for the delay of the coincidence logic, the cable length, the signal digitization time, and the muon time of flight.

A two-step procedure is performed to reconstruct the global muon track.
The first step consists of a local fit, performed separately in each chamber with at least 3 TDC signals, to solve the left/right ambiguity and identify a local track segment. A chi-square linear fit is performed on all possible combinations of hit position and lateralities. This iterative procedure does not rely on the trigger primitive information, as neither the filtering mask nor the lateralities assumed by the ANNs are provided.
The local segment is reconstructed as the combination with the best $\chi^2/ndof$, provided a minimum threshold of $\chi^2/ndof < 4$ is matched.
In the second step, a combined telescope-wide track is built using the 2 external $\phi$-view chambers and refitting their hits with the left/right ambiguities solved in the previous step to derive the global track parameters. In order to reject low-quality tracks, the minimum number of hits in the external chambers is required to be larger or equal to 7, and the $\chi^2/ndof$ to be less than 4. Additionally, the $\theta$-chamber should also have a reconstructed track in the orthogonal direction. This reconstruction method is thus completely independent from the triggering macro-cell TDC hits.
The angle of trigger primitives generated by the FW implementation of the algorithm is finally compared with the reconstructed angle of the global tracks, and the \tzero provided by the trigger primitive is compared with the coincidence time of the scintillator. 

\begin{figure}[htb]
  \centering
  \subfloat[]{\includegraphics[width=0.5\textwidth]{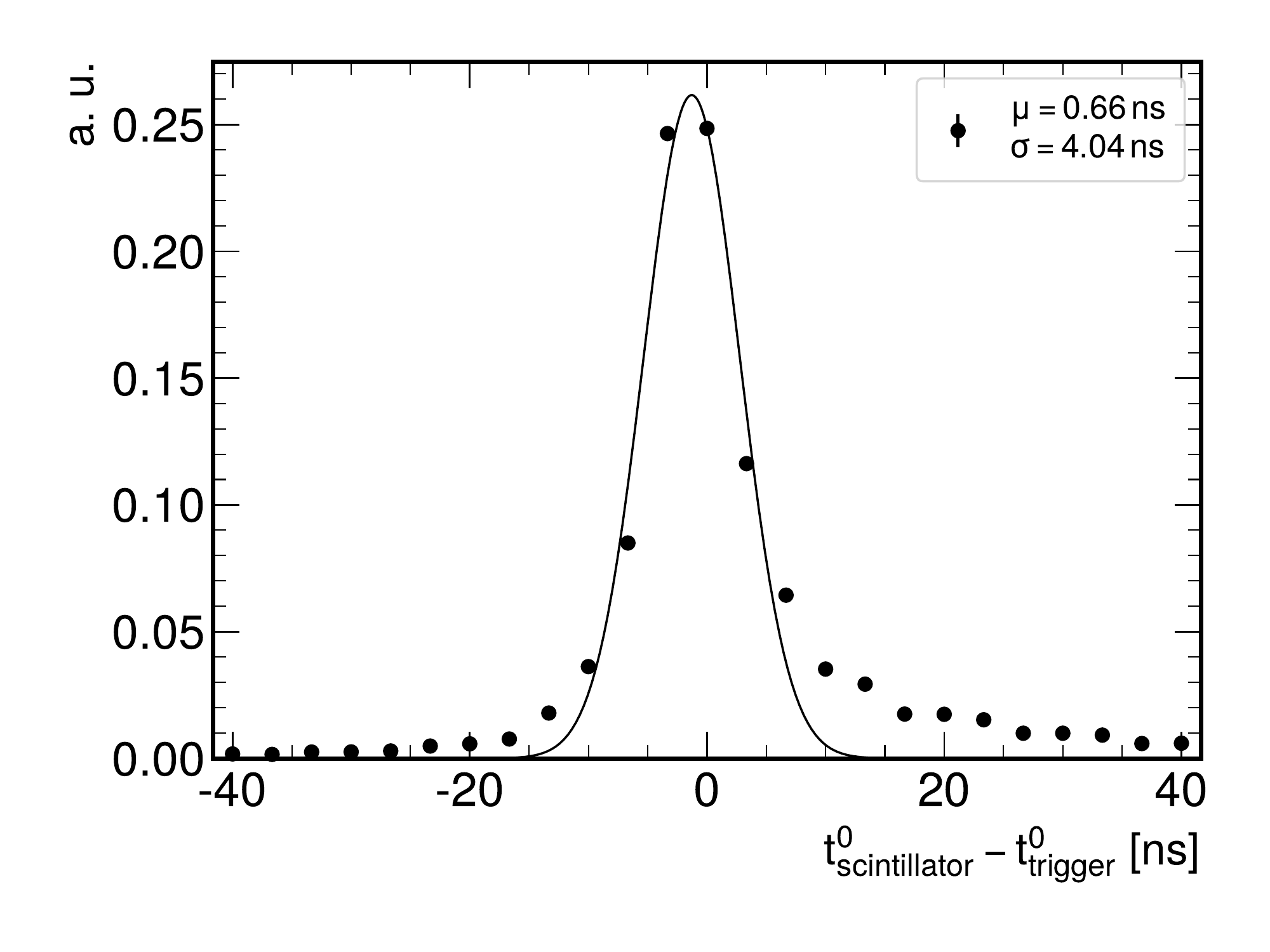}\label{fig:time_res_34_data}}
  \hfill
  \subfloat[]{\includegraphics[width=0.5\textwidth]{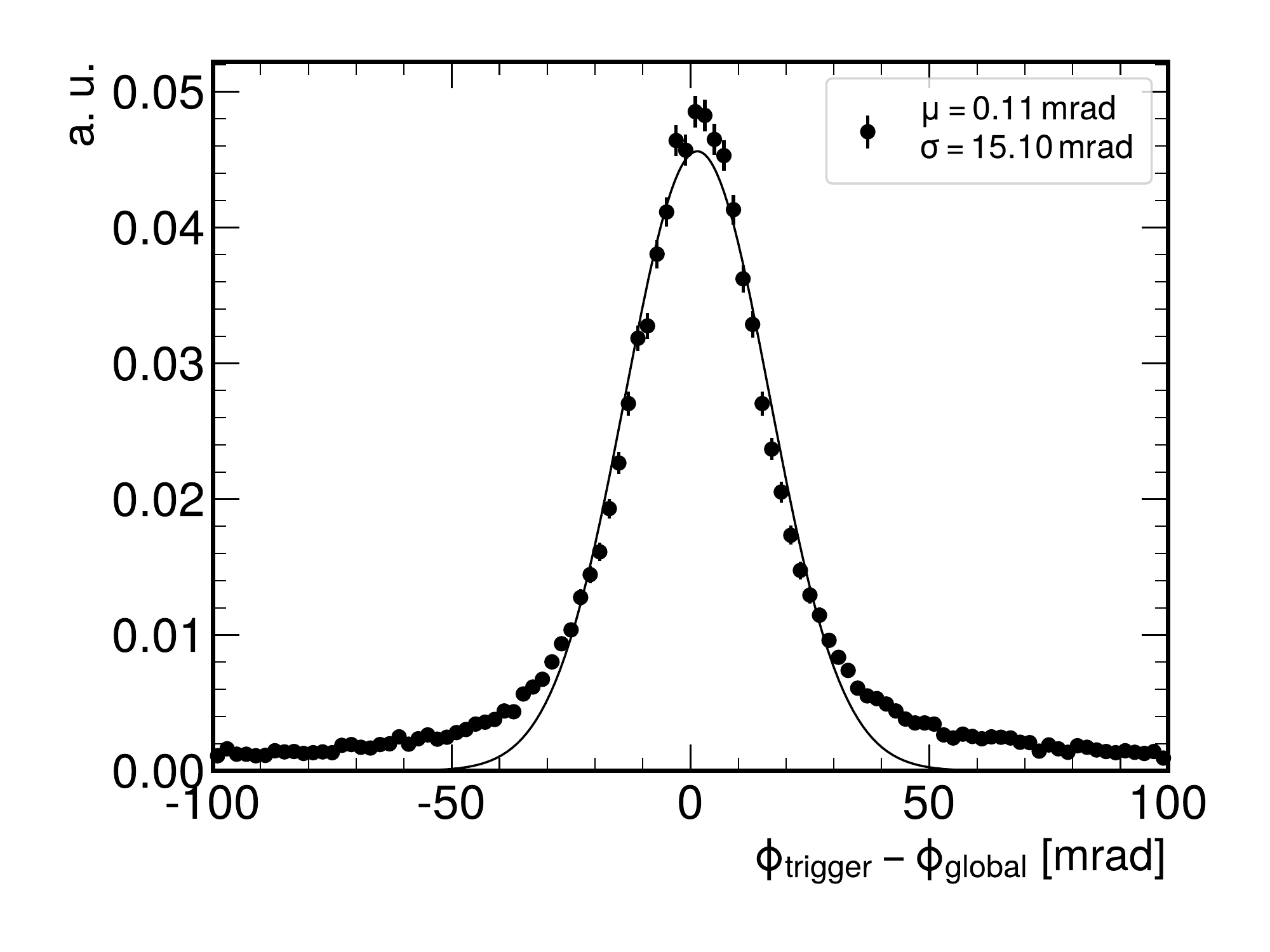}\label{fig:angle_res_34_data}}
  \caption{(a) Time resolution between the trigger \tzero and the external scintillator time. (b) Resolution of the trigger track angle with respect to the global track.}
\end{figure}

The estimated time resolution is obtained from a Gaussian fit to the bulk of the aforementioned time difference distribution, reported in Figure~\ref{fig:time_res_34_data}, resulting in $\sigma_t=4.1\,\ns$, in good agreement with the estimation provided by the emulator. 
Triplets and quadruplets are defined on data on the basis of the number of hits matched with the local segment reconstructed inside the macro-cell.
A fraction of approximately $11.5\,\%$ \tri are observed in the analyzed dataset over a total number of about $44,000$ reconstructed local segments.
Both the \qua and \tri time distributions show comparable bulk resolutions to the inclusive distribution, with $\sigma_t^\qua=4.0\,\ns$ and $\sigma_t^\tri=4.8\,\ns$ respectively, in agreement with the studies based on the emulator.

An asymmetric tail in the time difference distribution, affecting a fraction of less than $10\,\%$ of the events, can be observed in Figure~\ref{fig:time_res_34_data}, where both \tri and \quad contributes proportionally to their respective populations. 
This effect is not inherent to the trigger logic, as it is mainly due to a combination of ill-defined events and instrumental effects.
It should be mentioned that no corrections for the jitter induced by the scintillator readout is accounted for in this work, primarily focused on the proof-of-principle provided by the first implementation of the algorithm as a demonstrator of its strategy.

The angular resolution can be evaluated by comparing the trigger primitive angular parameter with either the local segment angle or the reconstructed global track angle.
While the former provides an intrinsic validation of the fit performed in the FW implementation of the algorithm, where in principle the same information is used in both the trigger and offline segment linear fit, the latter provides a mostly unbiased comparison as the offline global track is reconstructed from hits with a large lever arm across them and fully uncorrelated from those provided in input to the trigger.

Multiple scattering effects induced by the material of the mini-DT chambers are not a priori negligible, as the unfiltered cosmic rays' energy spectrum at sea level is particularly soft. However, the impact of multiple scattering on the $\phi$ angle reconstruction is expected to be small, and it is neglected in this study.

The resolution on the trigger primitive angular parameter with respect to the global track is measured as $\sigma_\phi=15.6\,\mrad$ overall, and reported in Figure~\ref{fig:angle_res_34_data}. The breakdown of the angular resolution for \qua and \tri signals results in $\sigma_\phi^\qua=15.1\,\mrad$ and $\sigma_\phi^\tri=25.9\,\mrad$, respectively.

\begin{figure}[htb]
  \centering
  \subfloat[]{\includegraphics[width=0.5\textwidth]{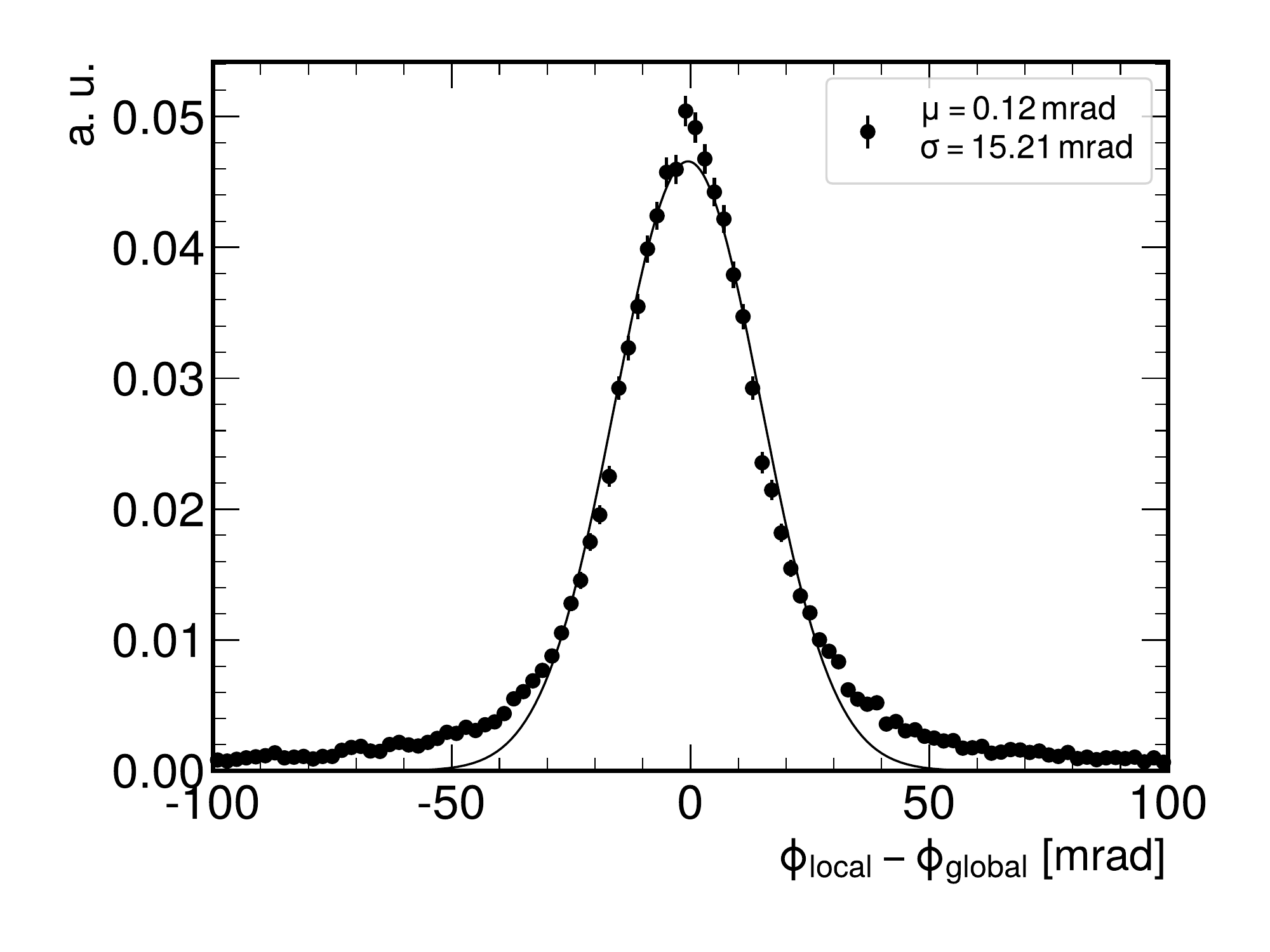}\label{fig:angle_res_seg_vs_track_data}}
  \hfill
  \subfloat[]{\includegraphics[width=0.5\textwidth]{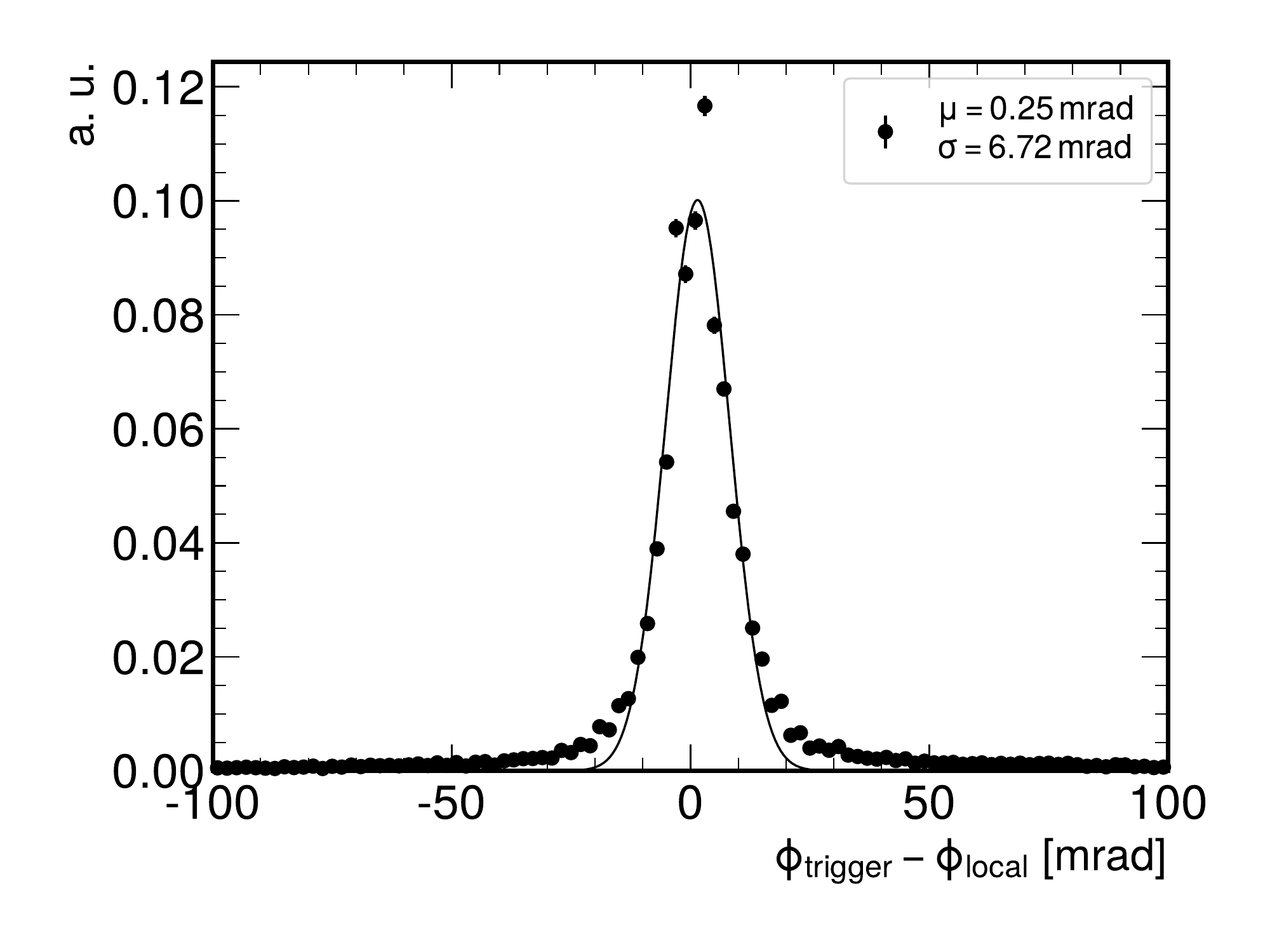}\label{fig:angle_res_vs_seg_data}}
  \caption{(a) Distribution of the difference in the angle between the global track and the local segment reconstructed in the triggering chamber. (b) Distribution of the difference in the angle between the trigger primitive and the local segment reconstructed in the macro-cell.}
\end{figure}

The estimate of the resolution based on the global track reconstruction is systematically limited by the intrinsic resolution of the telescope, which is significantly larger than the trigger angular resolution itself. The comparison of the angular parameter between the global tracks and the local segment reconstructed in the middle chamber (whose hits are not included in the global track) is described in Figure~\ref{fig:angle_res_seg_vs_track_data}. 

The limited angular resolution of the telescope can be evaded by comparing trigger primitive directly to the local segment, reconstructed using only the hits collected in the macro-cell. The result is shown in Figure~\ref{fig:angle_res_vs_seg_data}, and a standard deviation of $6\,\mrad$ is measured. Whilst this value cannot be directly interpreted as an intrinsic angular resolution, it provides a clear indication of the excellent trigger performance, which is comparable to the one of the offline reconstruction.

\begin{figure}[htb]
    \centering
    \includegraphics[width=0.495\textwidth]{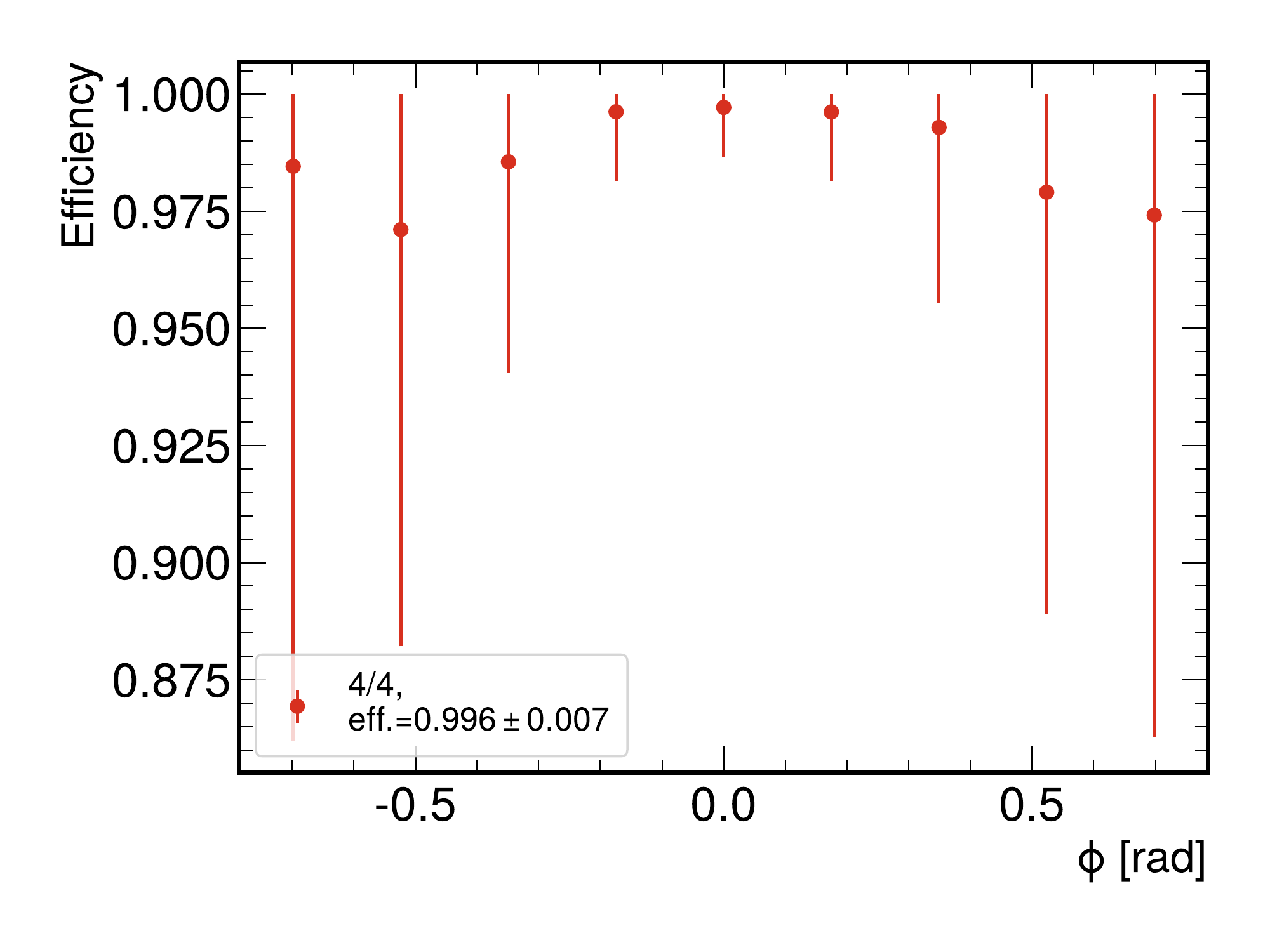}
    \caption{Trigger efficiency as a function of the track angle for \qua signals.}
    \label{fig:eff_vs_angle_data}
\end{figure}

The trigger efficiency is defined in data as the fraction of the reconstructed global tracks for which an associated trigger primitive has been generated by the algorithm implemented in FW. The overall measured efficiency is $\varepsilon=98.8 \pm 0.7\,\%$, while the \qua hits efficiency is $\varepsilon^{\qua}=99.6 \pm 0.7\,\%$, in excellent agreement with the expected results from the emulator studies. 
The dependency of the trigger efficiency on the particle's crossing angle is shown in Figure~\ref{fig:eff_vs_angle_data} for \qua tracks. 
The telescope and scintillator acceptances, combined with the natural angular spectrum of cosmic rays, provide a limiting factor at large angles where only a small number of tracks can be reconstructed.

\section{Concluding Remarks and Future Outlook}\label{sec:concl}

A novel approach to the generation of local trigger primitives for drift tubes muon detectors has been presented, based on an hybrid model integrating artificial neural networks and analytical methods.
The key feature of the proposed algorithm is the efficient use of resources to perform the identification and disambiguation of the signals pertaining to the expected signal thanks to two ANN layers.
Analytical relations are eventually implemented to extract the track parameters on the basis of the unique combination of viable signals identified by the ANNs, thus by-passing the otherwise unavoidable large level of combinatorial.
A demonstrator of the algorithm is deployed on state-of-the-art FPGA for a single macro-cell configuration. 
The performances of the algorithm are evaluated both on a private simulated sample and on data collected from cosmic rays with an experimental testbed, and found to be competitive with the figures expected from the offline reconstruction.
The footprint of the hardware implementation of the algorithm is measured. The pivotal components of the algorithm, the filtering and disambiguation ANNs block, are implemented in HLS thanks to the hls4ml library. The optimization of the ANNs through pruning and weight quantization allows for a total usage of 11k LUTs, accounting for less than $2\,\%$ of the available resources of the XCKU115 FPGA used in this study. No DSP is used for either ANN block. The total latency of the two ANNs blocks is measured in 5 clock cycles at $40\,\MHz$, and the execution of the entire algorithm can be fully pipelined.

An extension of the proposed algorithm is foreseen to generate trigger primitives from an entire DT detector geometry. An horizontal extension to cover a larger number of channels over the same chamber is under study. By replicating a series of macro-cell entities in an array, it is possible to span any 4-layer chamber configuration. An overlapping set of 4 channels across two consecutive macro-cells is expected to maximize the trigger primitive efficiency and acceptance. As the processing of all trigger blocks can be pipelined, it is expected to be able to funnel hits from multiple macro-cells into a smaller number of ANN blocks, multiplexing all FPGA stages down to a single (or few) track parameter estimation block for an entire macro-cell array.
The extension of the proposed logic toward the vertical integration, correlating signals across multiple chambers is also under study.

\bibliographystyle{ieeetr}
\bibliography{Bibliography}{}

\begin{thebibliography}{10}

\bibitem{cit:CMS}
T.~C. collaboration, ``{The CMS experiment at the CERN LHC. The Compact Muon
  Solenoid experiment},'' {\em JINST}, vol.~3, p.~S08004. 361 p, 2008.
\newblock Also published by CERN Geneva in 2010.

\bibitem{cit:CMS_muTDR}
J.~G. Layter, {\em {The CMS muon project: Technical Design Report}}.
\newblock Technical design report. CMS, Geneva: CERN, 1997.

\bibitem{cit:MAD}
F.~Gonella and M.~Pegoraro, ``{"The MAD", a Full Custom ASIC for the CMS Barrel
  Muon Chambers Front End Electronics},'' 2001.

\bibitem{cit:phase2Elec}
A.~Triossi {\em et~al.}, ``{Electronics Developments for Phase-2 Upgrade of CMS
  Drift Tubes},'' {\em PoS}, vol.~TWEPP2018, p.~035, 2019.

\bibitem{cit:gbt_FPGA}
S.~Baron, J.~P. Cachemiche, F.~Marin, P.~Moreira, and C.~Soos, ``{Implementing
  the GBT data transmission protocol in FPGAs},'' in {\em {Proceedings, Topical
  Workshop on Electronics for Particle Physics (TWEPP09)}}, CERN, CERN, 2009.

\bibitem{cit:ttc}
B.~Taylor, ``Ttc distribution for lhc detectors,'' {\em Nuclear Science, IEEE
  Transactions on}, vol.~45, pp.~821 -- 828, 07 1998.

\bibitem{cit:CMS_meantimer}
F.~Gasparini, R.~Giantin, R.~Martinelli, A.~Meneguzzo, G.~Pitacco, P.~Sartori,
  R.~Soggia, P.~Zotto, M.~Andlinger, F.~Szoncso, G.~Walzel, C.-E. Wulz,
  G.~Bencze, M.~{Della Negra}, D.~Peach, E.~Radermacher, C.~Seez, and
  G.~Wrochna, ``Bunch crossing identification at lhc using a mean-timer
  technique,'' {\em Nuclear Instruments and Methods in Physics Research Section
  A: Accelerators, Spectrometers, Detectors and Associated Equipment},
  vol.~336, no.~1, pp.~91--97, 1993.

\bibitem{Puerta-Pelayo:904792}
J.~Puerta-Pelayo, M.~C. Fouz-Iglesias, and P.~García-Abia, ``{Parametrization
  of the Response of the Muon Barrel Drift Tubes},'' tech. rep., CERN, Geneva,
  Oct 2005.
\newblock This note is a revised version of CMS-IN-2005-037.

\bibitem{Benettoni:2007zz}
M.~Benettoni, F.~Gasparini, F.~Gonella, A.~Meneguzzo, S.~Vanini, G.~Zumerle,
  and G.~Bonomi, ``{CMS DT chambers: Optimized measurement of cosmic rays
  crossing time in absence of magnetic field},'' 10 2007.

\bibitem{cit:CMS_performance}
S.~Chatrchyan {\em et~al.}, ``{The Performance of the CMS Muon Detector in
  Proton-Proton Collisions at $\sqrt{s}$ = 7 TeV at the LHC},'' {\em JINST},
  vol.~8, p.~P11002, 2013.

\bibitem{Coelho:2020zfu}
C.~N. Coelho, A.~Kuusela, S.~Li, H.~Zhuang, T.~Aarrestad, V.~Loncar,
  J.~Ngadiuba, M.~Pierini, A.~A. Pol, and S.~Summers, ``{Automatic deep
  heterogeneous quantization of Deep Neural Networks for ultra low-area,
  low-latency inference on the edge at particle colliders},'' 6 2020.

\bibitem{chollet2015keras}
F.~Chollet {\em et~al.}, ``Keras.'' \url{https://keras.io}, 2015.

\bibitem{2017arXiv171001878Z}
M.~{Zhu} and S.~{Gupta}, ``{To prune, or not to prune: exploring the efficacy
  of pruning for model compression},'' {\em arXiv e-prints},
  p.~arXiv:1710.01878, Oct. 2017.

\bibitem{Duarte:2018ite}
J.~Duarte {\em et~al.}, ``{Fast inference of deep neural networks in FPGAs for
  particle physics},'' {\em JINST}, vol.~13, no.~07, p.~P07027, 2018.

\end{thebibliography}
 
\end{document}